\documentclass[aps,twocolumn,prl,groupedaddress,showkeys]{revtex4-1}
\usepackage{graphicx}
\usepackage{booktabs}
\usepackage[utf8]{inputenc}
\usepackage{wrapfig}
\begin{document}
\title{Light-induced \textit{breathing} in photochromic yttrium oxy-hydrides} 

%
%

\author{Elbruz Murat~Baba\textsuperscript{1,4}, Jose~Montero\textsuperscript{2,$\star$} , Evgenii~Strugovshchikov\textsuperscript{3}, Esra {\"{O}}zkan~Zayim\textsuperscript{4,5}\ and Smagul~Karazhanov\textsuperscript{1}~\\[-3pt]\normalsize\normalfont ~\\
\textsuperscript{1}{Department for Solar Energy\unskip, Institute for Energy Technology\unskip, NO-2027\unskip, Kjeller\unskip, Norway}~\\
\textsuperscript{2}{Department of Engineering Sciences\unskip, The {\AA}ngstr{\"{o}}m Laboratory\unskip, Uppsala University\unskip, Uppsala\unskip, SE-75121\unskip, Sweden}~\\
\textsuperscript{3}{Institute of Physics\unskip, University of Tartu\unskip,  Tartu\unskip, Estonia}~\\
\textsuperscript{4}{Nano Science \& Nano Engineering Department\unskip, Istanbul Technical University\unskip, Istanbul\unskip, 34469\unskip, Turkey}~\\
\textsuperscript{5}{Physics Department\unskip, Istanbul Technical University Faculty of Science and Letters\unskip, Istanbul\unskip, 34469\unskip, Turkey\newline \textsuperscript{$\star$} email : jose.montero-amenedo@angstrom.uu.se (Jose~Montero), }}


\date{\today}

\begin{abstract}

When exposed to air, metallic yttrium dihydride YH\ensuremath{_{2}} films turn into insulating and transparent yttrium oxy-hydride (YHO). The incorporation of oxygen causes the lattice expansion of YH\ensuremath{_{2\ }} and the emergence of photochromic properties. However, the oxidation of YH\ensuremath{_{2}} is not completely irreversible: under illumination some oxygen atoms move towards the YHO surface, leaving behind an oxygen-deficient bulk (responsible for the photochromic darkening and observed lattice contraction). Under illumination, and according to experimental evidence, some oxygen atoms can effectively leave the film, being replaced by other oxygen atoms once the illumination has stopped, \textit{i.e.}, YHO `breathes' when subjected to illumination/darkness cycling. Based on this `breathing', YHO films become more hydrophobic under illumination conditions than when kept in darkness. \emph{Ab-initio} calculations point to the light-induced weakening of the Y-O bond as the possible mechanism for explaining these experimental observations. 

\end{abstract}.  

\maketitle 


\section{Introduction}

 Yttrium hydride and other rare-earth  hydrides are extremely reducing agents, a feature that complicates considerably their study. For their adequate handling in air, rare-earth  hydride thin films are usually protected against oxidation by, for example, Pd capping layers \unskip~\cite{388382:8582484}. However, the incorporation of oxygen in rare-earth hydrides after intentional exposure to air \unskip~\cite{388382:8582534,388382:8582572,388382:8582483}, or even through accidental contamination \unskip~\cite{388382:8582482}, leads to the formation of  oxy-hydrides, which exhibit very interesting properties. One of the pioneer works on this family of materials was carried out by Miniotas \emph{et al.}\unskip~\cite{388382:8582482}, who reported gigantic electrical resistivity in oxygen-containing gadolinium hydride. Later, Mongstad \emph{et al.} \unskip~\cite{388382:8582563} reported photochromic properties in oxygen-containing yttrium hydride, a feature observed very recently by Nafezarefi \emph{et al.} \unskip~\cite{388382:8582534} in other rare-earth oxy-hydrides such as dysprosium, gadolinium or erbium oxy-hydrides. The photochromism in Y-related compounds can be traced back to Ohmura \emph{et. al} \unskip~\cite{388382:8582546}, who observed light-induced reversible darkening in yttrium hydride thin films subjected to high pressures ($\sim$ GPa). Despite the importance of the discovery, the emergence of this new inorganic photochromic material went unnoticed at that time, presumably because the pressure range required is not suitable for practical applications. Today, however, it is known that yttrium oxy-hydride --hereinafter referred in the text simply as YHO, a notation that, in principle, is not related to the stoichiometry of the compound, which will be discussed later-- as well as other rare-earth oxy-hydrides, are photochromic at room temperature and at ambient pressure; hence, YHO, as an inorganic photochromic material has multitude of potential applications \unskip~\cite{388382:8581903}. However, the origin of the photochromic mechanism in YHO is still open to debate \unskip~\cite{388382:8582573}, and in this sense, the wettability of the YHO surface under illumination and darkness conditions has been found to provide valuable insights: the present work reports on the light-induced hydrophobicity enhancement in YHO, \textit{i.e.}, the reduction of the surface energy under illumination. All oxides and nitrides of low-electronegativity metals can exhibit hydrophobicity \unskip~\cite{388382:8582513,388382:8582588}. It can be, therefore, expected that YHO exhibits hydrophobic properties as well. However, while the surface of yttrium oxy-hydride increases its hydrophobicity when illuminated, other metal oxides turn into hydrophilic under UV illumination. In the latter case, the formation of electron-hole pairs under illumination leads to the creation of defect sites, where hydroxyl groups can be adsorbed, leading to hydrophilic properties \unskip~\cite{388382:8582538}. Generally, when metal oxides are stored in darkness during periods of time ranging from 7 to 50 days \unskip~\cite{388382:8582580,388382:8582537}, oxygen replaces back the adsorbed hydroxyl groups, giving raise to hydrophobicity. In the present work, the unexpected behavior observed in YHO, \textit{i.e}., the enhancement of the hydrophobic properties under illumination, has been found to be caused by the same reason, that is, the oxygen-enrichment of the surface under illumination. This behavior provides new insights on the photochromic mechanism in yttrium oxy-hydride: under illumination, oxygen atoms are released from the lattice (in consequence the YHO lattice contracts); some of these O atoms can reach the surface, causing the enhancement of the hydrophobic properties. The displaced oxygen atoms leave behind an oxygen-deficient structure responsible for the optical darkening of the film. In darkness, the YHO lattice expands back as a consequence of the filling of the oxygen vacancies by oxygen atoms, allowing the film to bleach back to its original state. Since YHO expands/contracts reversibly under dark/illumination cycling (produced by the displacement inwards/outwards of oxygen atoms) we refer to this process as \textit{breathing}. The \textit{breathing} hypothesis presented in this work explains the photochromic behavior as well as the wettability change in YHO and it is supported by both theoretical and experimental results.
\section{Methods}

Oxygen containing yttrium hydride thin films were prepared onto glass substrates following a two-step deposition process consisting on the fabrication by magnetron sputtering of YH\ensuremath{_{2}} metallic films (in a Leybold Optics A550V7 sputter unit) followed by a post-deposition oxidation process in air. Further details on the synthesis process of photochromic YHO can be found elsewhere \unskip~\cite{388382:8582572,388382:8582486}. A cold white LED array from Thorlabs (colour temperature 4600-9000 K) was used as illumination source for the photo-darkening experiments.The crystallographic structure of the obtained films was characterized by using x-ray diffraction (XRD) in a Bruker Siemens D500 spectrophotometer (CuK\textit{\ensuremath{\alpha } }radiation, parallel beam geometry). The composition and surface oxidation states were studied by x-ray photoelectron spectroscopy (XPS) in an Ulvac PHI Quantera II instrument. Surface roughness characterizations were performed using Atomic Force Microscopy with area of 5 $\mu$m${^{2}}$ from Photonic-tech Picostation. The optical transmittance (\textit{T}) of the YHO films in the clear and photodarkened state was measured using an Ocean Optics spectrophotometer QE65000 and a Perkin-Elmer Lambda-900 with integrating sphere. Contact angle (CA) measurements were performed using KSV Attension Optical Tensiometer under air. 5 $\mu$l drop volume was used for each CA measurement and three different sessile droplets were measured on several substrates for each value and averaged with a standard deviation of \ensuremath{\pm}2. CA values in the equilibrium ($\theta_e$) for water, ethylene glycol (EG), and methylene iodide (MeI)--both EG and MeI from Sigma-Aldrich--were used to calculate surface free energies of Yttrium oxy-hydride films at clear and photo-darkened state using the van Oss-Good-Chaudhury method\unskip~\cite{388382:8582545,388382:8582531}.

The calculations were performed with the Vienna Ab initio Simulation Package (VASP) code \unskip~\cite{388382:8581928,388382:8581929,388382:8581930}, based on density functional theory (DFT) using a plane-wave pseudopotential method together with the potential projector augmented-wave (PAW) \unskip~\cite{388382:8582478,388382:8582477,388382:8581933}. The generalized gradient approximation (GGA) in the scheme of Perdew-Burke-Ernzerhof (PBE) is used to describe the exchange-correlation functional \unskip~\cite{388382:8581930}. To describe the electron-ion interaction standard PAW-PBE pseudopotentials \unskip~\cite{388382:8582475} are used with 1\textit{s}\ensuremath{^{1}} for H, 2\textit{s}\ensuremath{^{2}}2\textit{p}\ensuremath{^{4}} for O and 4\textit{s}\ensuremath{^{2}}4\textit{p}\ensuremath{^{6}}4\textit{d}\ensuremath{^{1}}5\textit{s}\ensuremath{^{2}} for Y atoms as the valence-electron configuration. The plane wave functions of valence electrons are expanded in a plane wave basis set, and the use of PAW pseudopotentials allows a plane wave energy cutoff ($E_{cut}$). Only plane waves with kinetic energies smaller than $E_{cut}$ are used in the expansion. Reciprocal-space integration over the Brillouin zone is approximated through a careful sampling at finite number of \textbf{\textit{k}}-points using a Monkhorst-Pack mesh \unskip~\cite{388382:8581933}. We choose the energy cutoff to be 700 eV, and the Brillouin-zone sampling mesh parameters for the \textbf{\textit{k}}-points set are 8\ensuremath{\times}8\ensuremath{\times}8. In the optimisation process the energy change is set to $1\times10^{-6}$ eV. The charge densities are converged to $1\times10^{-6}$ eV in the self-consistent calculation. The range-separated hybrid Heyd-Scuseria-Ernzerhof (HSE06) functional is used for density of states calculations \unskip~\cite{388382:8581935,388382:8582473,388382:8582472}. The hybrid functional requires a standard value of the (short-range) Hartree-Fock exchange (21\%) mixed with a portion of PBE exchange (79\%), also known as the HSE06 hybrid functional \unskip~\cite{388382:8582473,388382:8582472}. Selection of the parameter has been performed as an inverse value of infinity dielectric constant that is valid, if the energy band gap of these systems is larger than 3 eV. 

\section{Results}

\subsection{Hydrophobicity control through light illumination}
\begin{figure*} 
\includegraphics[width=0.99\textwidth]{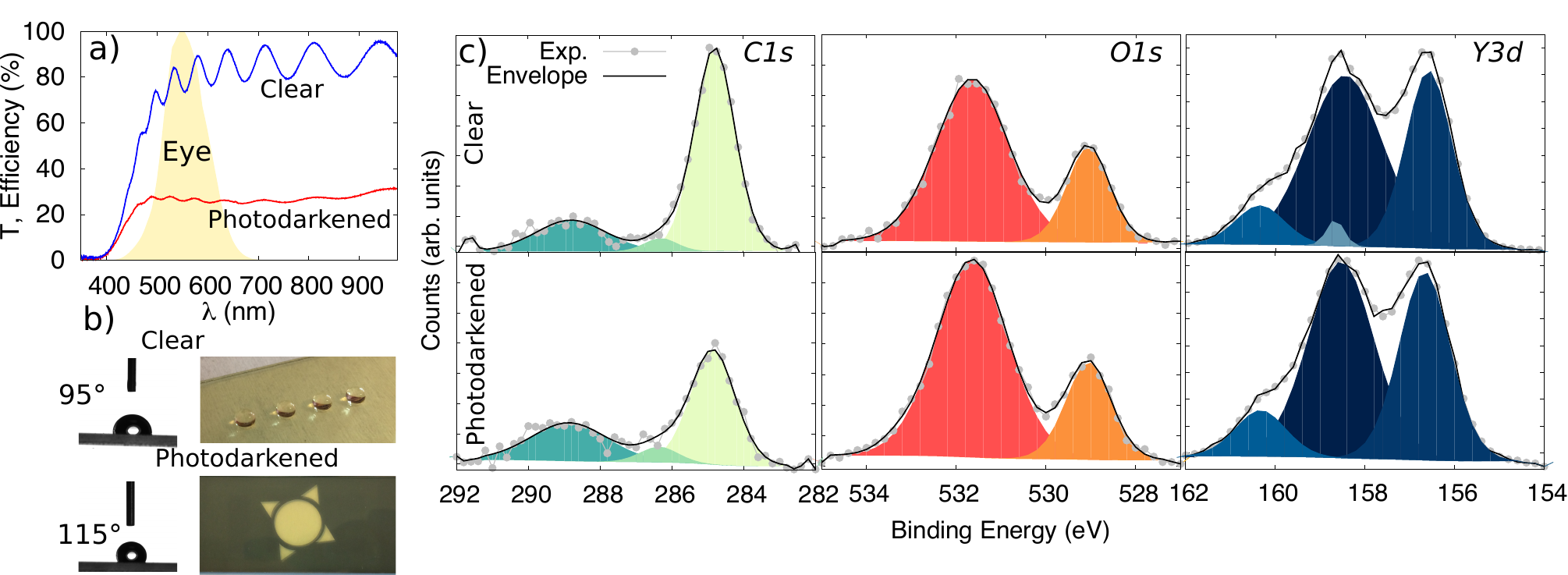}
\caption{Different properties measured in the clear state and photodarkened state (after illumination) for a 1400 nm-thick YHO film: transmittance compared to the luminous efficiency of the human eye in photopic vision (panel a), contact angle photographs for water, a photo of water droplets onto a clear film, as well as the same film after being illuminated  under a sun-shaped mask during 1 h (panel b). XPS spectra corresponding to C1\emph{s}, O1\emph{s} and Y3\emph{d} (panel c). }
\label{fig1}
\end{figure*}

\begin{table*}
\centering
\begin{center} 
\begin{tabular}{ccccccccc}

& \hspace{0.1in}$\theta_e$(Water)\hspace{0.1in} & \hspace{0.1in}$\theta_e$(EG)\hspace{0.1in}  & \hspace{0.1in} $\theta_e$(MeI)\hspace{0.1in}  & $\gamma_{total}$  & $\gamma ^{LW}$  & $\gamma ^{AB}$ & $\gamma^{+}$  & $\gamma^{-}$ \\ 
   
& ($^\circ$) &   ($^\circ$) &  ($^\circ$) & \hspace{0.1in}(mJ/m$^2$)\hspace{0.1in} &  (mJ/m$^2$) \hspace{0.1in}&  \hspace{0.1in}(mJ/m$^2$)\hspace{0.1in} &  \hspace{0.1in}(mJ/m$^2$)\hspace{0.1in} &(mJ/m$^2$) \\ 
\hline 
\hline
Clear state & 95 & 81 & 43 & 46.55 & 38.07 & 8.98 & 2.98 & 6.76  \\
\hline
Photo-darkened State & 115 & 81 & 60 & 28.88 & 28.58 & 0.31 & 0.18 & 0.13 \\
\hline
\end{tabular}
\caption{Equilibrium contact angle values ($\theta_e$) for water, ethylene glycol (EG) and methylene iodine (MeI) as well as the measured total  surface energy ($\gamma_{total}$) and its components: Liftshitz-van del Waals interactions term ($\gamma^{LW}$) and acid-base interaction term ($\gamma^{AB}$)--calculated from the Lewis acid and base parameters ($\gamma^+$ and $\gamma^-$ respectively)--. All data given for the clear and photo-darkened state.\label{table_wet} }
\end{center}
\end{table*}

\begin{table}
\centering
\begin{center} 
\begin{tabular}{ccc}
\toprule
   & Clear State \hspace{0.1in} & Photo-darkened State   \\ 
     & \%  & \%  \\ 
\hline
Carbon & 48 & 36  \\

Oxygen & 36 & 48 \\

Yttrium & 16 & 16 \\
\hline
\end{tabular}

\caption{XPS measurement for oxygen, yttrium and carbon content change at the yttrium oxy-hydride coating surface before and after illumination.\label{table_comp} }
\end{center}

\end{table}

YHO thin films exhibit photochromic properties, that is, YHO films undergo a reversible decrease of their optical transmittance when illuminated with light of adequate energy and intensity \unskip~\cite{388382:8582486}. Figure \ref{fig1} (a) shows the transmittance in the clear and photodarkened state for 1400 nm-thick YHO film. This film decreased its luminous transmittance, $T_{lum}$ from 78.5\% to 26.7\% after illumination (for a definition of $T_{lum}$ and how to calculate it, visit ref. \unskip~\cite{388382:8582468}). The luminous efficiency of the human eye (photopic vision) is presented in Figure \ref{fig1} (a) for comparison \unskip~\cite{388382:8582468}. How to obtain such optical contrast by illumination will be discussed in detail in next section.

Non-illuminated (clear) YHO thin films show hydrophobicity with equilibrium contact angle ($\theta_e$) values of 95$^\circ$ for water (see Table \ref{table_wet}); however, $\theta_e$ values increased to 115$^\circ$  after illumination (again in the case of water), see Table \ref{table_wet} and Figure \ref{fig1} (b). Table \ref{table_wet} also shows $\theta_e$ for ethylene glycol (EG) and  methylene iodine (MeI) for the clear and photo-darkened  states. In the case of MeI, $\theta_e$ also increases after illumination, from 43$^\circ$ to 60$^\circ$, while remain constant for EG. AFM studies performed in such films revealed a relatively smooth surface, being the RMS value of around 8 nm.

The observed initial hydrophobicity of the YHO films (clear state) can be explained by the electronic structure of rare earth elements: according to a detailed experimental analysis of the entire rare earth oxide series carried out by Azimi et al. \unskip~\cite{388382:8582588}, the unfilled 4\textit{f} orbitals shielded by a full octet of electrons (from the 5\textit{s}\ensuremath{^{2}}\textit{p}\ensuremath{^{6}} shell) which is characteristic of rare earth oxides, results in a lower tendency of such compounds to form hydrogen bonds with the adjacent water molecules \unskip~\cite{388382:8582588,388382:8582533}. However, hydrophobicity is not exclusive of the lanthanide \textit{f}-shell group, but it can be achievable in any metal oxide provided that the electronegativity of such metal is low enough \unskip~\cite{388382:8582513}. The low electronegativity of Y and the fact that the surface is composed mainly by yttrium oxide \unskip~\cite{388382:8582584}, explains the high $\theta_e$ shown in  Table \ref{table_wet}. 

Under illumination, a decrease of the hydrophobicity, caused by the creation of electron-hole pairs, would be expected, as observed in other metal oxides \unskip~\cite{388382:8582538,388382:8582580,388382:8582536,388382:8582579,388382:8582578,388382:8582518}. However, this is not the tendency observed in YHO, in which, as described before, the hydrophobicity is enhanced under illumination. In this case, the observed light-induced decrease of wettability has to be explained through changes in\textit{} surface composition, namely the oxygen-to-metal ratio. In metal oxides, coordinatively unsaturated oxygen atoms work as a Lewis base while the metal cations work as a Lewis acid. Combined Lewis acid and base orientation of the surface causes high affinity towards water molecules \unskip~\cite{388382:8581940} and, therefore, the oxygen-to-metal ratio in the surface is crucial for understanding the wettability properties \unskip~\cite{388382:8582577}. 

In order to study the compositional changes of the YHO surface, XPS measurements were performed before and after illumination and the results presented for C1\textit{s}, O1\textit{s} and Y3\textit{d} in Figure \ref{fig1} (c).  See Table \ref{table_comp} for the quantification of the different elements by XPS. The carbon (adventitious) C1\textit{s} signal can be deconvoluted into three different contributions; the signal corresponding to C-C has been established at 284.8 eV as a charge correction reference. Other contributions are C-O-C at 286.3 eV and O-C=O at 288.8 eV \cite{Beamson93}. After illumination, the carbon content on the surface decreases, see Table \ref{table_comp}. This decrease is accused mainly by the C-C contribution, Figure \ref{fig1} (c). Since the content of C in the surface decreases, the increase of adsorbed hydrocarbons  is ruled out, in our case, as the possible cause for the light-induced enhancement of the hydrophobicity \unskip~\cite{388382:8582495,388382:8582512,388382:8582551}.

The O1\textit{s} signal is composed by two contributions at 529.0 eV and 531.2 eV; the former can be attributed to O atoms bound to Y atoms, whereas the latter can be assigned to physisorbed O \cite{Craciun1999}. After illumination, the O content of the surface increases, see Table \ref{table_comp}. This increase takes place  both for  O bound to Y as well as for physisorbed O. However, the  largest increase is observed in the later, Figure \ref{fig1} (c).  

The obtained results for Y3\textit{d} correspond very well to the Y$_2$O$_3$ stoichiometry, \textit{i.e.}, at the top surface consist of Y$_2$O$_3$, in agreement with our previous studies \unskip~\cite{388382:8582584}. The Y3\textit{d} has well resolved spin-orbit components, namely Y3\textit{d}$_{3/2}$ and Y3\textit{d}$_{5/2}$. These components can be deconvoluted into yttrium carbonates, with small contributions at 158.6 eV and 160.3 eV, as well as the main Y$_2$O$_3$ contribution at 156.6 eV and 158.4 eV, Figure \ref{fig1} (c) \cite{Craciun1999}. Not surprisingly there is no change in Y content during illumination, but the carbonate component does decrease in favour of the Y$_2$O$_3$ component.

Surface energy calculations, performed using the Oss-Chaudhury-Good method \unskip~\cite{388382:8582545,388382:8582531}, confirm the lower wettability through  reduction (under illumination) of the total surface energy ($\gamma_{total}$), see Table \ref{table_wet}. The enrichment in oxygen of the surface, confirmed by XPS, reduces the Lewis sites as the surface approaches the Y$_2$O$_3$ stoichometry. The  nonpolar Liftshitz-van der Waals surface energy component, $\gamma^{LW}$, also decreases from 38.07 mJ/m$^{2}$ to 28.57 mJ/m$^{2}$ while the  polar acid-base component, $\gamma^{AB}$ [where $\gamma^{AB}=2(\gamma^{+}\gamma^{-})^{1/2}$, being $\gamma^{+}$ and $\gamma^{-}$  the Lewis acid and base parameters of surface tension, respectively], decreases from 8.98 mJ/m\ensuremath{^{2}} to 0.31 mJ/m\ensuremath{^{2}} after illumination. 
\begin{figure*} 
\includegraphics[width=0.99\textwidth]{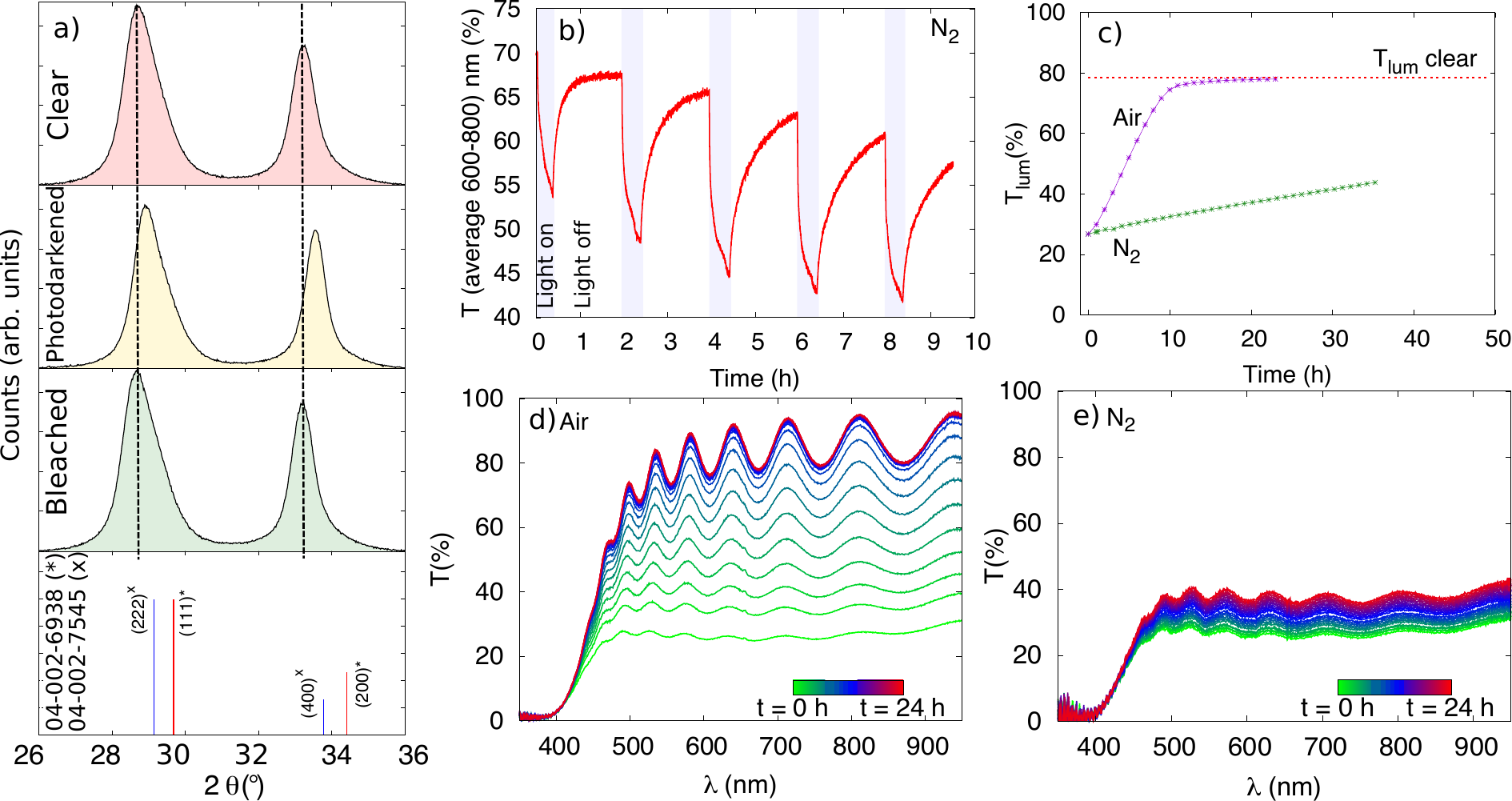}
\caption{X-ray diffraction patterns for the for a 1400 nm-thick YHO sample at the initial, dark and bleached state (panel a).  Average transmittance measured between 600-800 nm during 0.5 h illumination followed by 1 h darkness in a sample kept in N$_2$ atmosphere (panel b). Differences in the recovery dynamics for the sample in kept in air vs. the sample kept in N$_2$ atmosphere as illustrated by  the measurement of the luminous transmittance, Tlum, vs. time (panel c) and by a series of transmittance measurements vs. time (panels d and e). }
\label{fig2}
\end{figure*}
Hydrophobic yttrium-based oxides have been reported in the past \unskip~\cite{388382:8582513,388382:8582575}; in this case as the Y$_2$O$_{3-x}$ coatings approached the Y$_2$O$_{3}$ stoichiometry,
they showed increased contact angles\unskip~\cite{388382:8582575} which is consistent with metal-to-oxygen ratio of surface studies \unskip~\cite{388382:8582577}. 
Consequently, the enrichment in oxygen of the surface under illumination causes the light-induced hydrophobicity enhancement observed in YHO thin films. In the next section, the exchange of oxygen atoms between the film and the atmosphere, induced by illumination, is demonstrated.

\subsection{Light-induced ``breathing"}

As described above, photochromic yttrium oxy-hydride can be obtained by the oxidation in air of reactively sputtered metallic YH\ensuremath{_{2}} thin films.  The incorporation of oxygen in the YH\ensuremath{_{2}} lattice causes the increase of the lattice constant from $\sim$ 5.20 to $\sim$ 5.34 {\AA} \cite{MONGSTAD2011S812,388382:8582534,388382:8582572} and hence the displacement of the diffraction peaks towards lower angles. However under illumination, the lattice of the YHO films contracts back, but without reaching the original oxygen-free YH\ensuremath{_{2}} lattice constant \cite{MAEHLEN2013S119}: Figure \ref{fig2} (a) shows grazing incidence XRD patterns corresponding to an yttrium oxy-hydride sample in its initial (clear), illuminated (photodarkened) and recovered (bleached) states. The standard diffraction peaks for YH\ensuremath{_{2\ }} and Y$_2$O$_3$ according to the Joint Committee of Powder Diffraction Standards (JCPDS) card nums. 04-002-6938 and 04-002-7545  are also shown for comparison. 

 The analysis of the XRD  patterns revealed how the  films undergo an \textit{accordion-like} transformation, \textit{i.e.},  the YHO lattice contracts and expands when subjected to illumination/darkness cycles, respectively. In addition to these observations, our previous studies performed from a purely optical perspective pointed to the formation, under illumination, oxygen-deficient YHO domains within the YHO lattice \unskip~\cite{388382:8582573}. The filling factor (\textit{ff}) of such domains was predicted to be very small \unskip~\cite{388382:8582573} although it increased with illumination time\unskip~\cite{388382:8582573}.  The fact that the metallic domains are diluted in the dielectric YHO structure (Maxwell-Garnett effective medium approximation holds) explains why the darkened samples exhibit higher optical absorption than high optical reflectance\unskip~\cite{388382:8582563}.  In view of the above we postulate that, in YHO samples under illumination, some of the oxygen atoms are pushed towards the surface, leaving behind oxygen-deficient (metallic) domains which are responsible for the overall darkening of the film; in other words, YHO can be partially reverted to oxygen-defficient YHO by illumination.  Therefore, the photochromic change is caused by very subtle changes in oxygen content--as predicted by to our effective medium calculations \unskip~\cite{388382:8582573}--presumably difficult to detect experimentally. However, if this hypothesis is correct, after long enough illumination/darkness cycling in an oxygen-free atmosphere,  a small amount of oxygen atoms could effectively leave the YHO film every cycle, affecting the photochromic recovery; this is indeed the case as demonstrated by the following experiment: YHO thin films were subjected to 2 h period cycles (0.5 h illumination followed by 1.5 h darkness) inside a glove box filled with N\ensuremath{_{2}}. The O\ensuremath{_{2}} and H\ensuremath{_{2}}O content within the glove box was below  0.1  and 1.4 ppm, respectively.
  The average transmittance of the film  was measured between 600 and 800 nm during cycling and plotted in Figure \ref{fig2} (b). In the absence of air, the films lost part of their initial transparency in each cycle, not being able to recover fully.  After 4 weeks of continuous cycling within the glove box, the luminous transmittance $T_{lum}$ of the samples decreased from 78.5 \% in the non-illuminated state, to 26.7\%.  This heavily photodarkened films were allowed to bleach in total darkness, both in air and in N$_2$ atmosphere (glove box); the evolution of T$_{lum}$ is presented in Figure \ref{fig2} (c). It is evident from this figure that the bleaching speed in darkness of the photodarkened samples  was much slower inside the globe box than in air. In addition, a series of transmittance measurements have been performed during the recovery stage along a lapse of 24 hours both, in air  Figure \ref{fig2} (d) and  inside the glove-box (N$_2$ atmosphere), Figure \ref{fig2} (d). Note that, in both cases, the films were kept in total darkness (except for illumination with the spectrophotometer probe:  1 s each hour for performing the  measurements). The films kept in air recover totally their initial transparency after few hours (T$_{lum} clear$, presented in Figure \ref{fig2} (c) as an horizontal dashed line) while the films in N$_2$ recovered very little in the same period of time. Since there are not significant differences between the temperature inside and outside of the glove-box (both at \ensuremath{\sim } 20\ensuremath{^\circ} C), the data presented in Figure \ref{fig2} (c, d and e) strongly indicates that a source oxygen from the ambient is crucial for the adequate recovering of the photo-darkened films. This is consistent with the de-oxygenation hypothesis: a source oxygen is necessary in order to replace the oxygen atoms released during illumination; this points to a light-induced exchange of oxygen atoms between the film and the  environment.

Considering the low electronegativity of Y, the idea of oxygen being pushed out of the YHO lattice by illumination may look counterintuitive at first. However, it is important to consider that for achieving a noticeable darkening only very small amounts of oxygen are required to leave the film. Indeed, a \textit{ff } of just 2 \% of oxygen-deficient  YHO is known to cause a drop of the visible transmittance larger than 30\% \unskip~\cite{388382:8582486}. A theoretical model for understanding the light-induced oxygen release in YHO films is presented below.

\subsection{Theoretical considerations}
\begin{figure*} 
\includegraphics[width=0.99\textwidth]{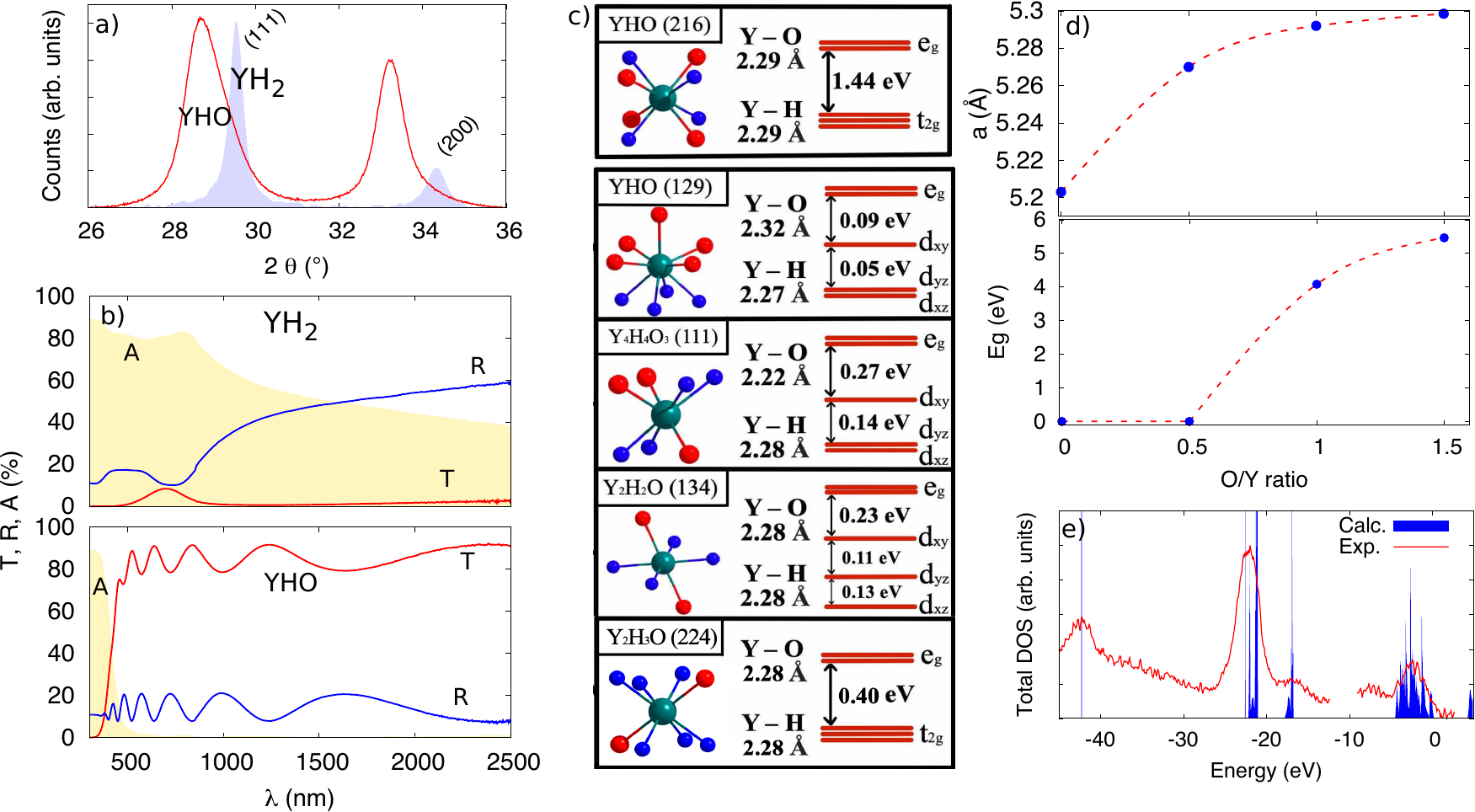}
\caption{XRD diffraction patters (panel a) as well as optical properties (panel b), namely transmittance (\emph{T}), reflectance (\emph{R}) and absorbance (\emph{A}) corresponding to YH$_2$ and compared to photochromic yttrium oxy-hydride (YHO). Schematic presentation of the bond coordination around Y, Y-O and Y-H bond lengths, and splitting of the Y 3$d$ states at the conduction band minimum for  YHO of different stoichiometries, YH$_x$O$_y$, where yttrium, hydrogen and oxygen atoms are presented as green, blue and red spheres, respectively (panel c).  Calculated lattice constant and bandgap (panel d)  as a function of O/Y ratio in the YH$_x$O$_y$ system.  Total density of states from \emph{ab initio} calculations as compared to experimental results from XPS measurements (panel e). }
\label{fig3}
\end{figure*}
Experimental evidence, presented above, points to light-induced oxygen exchange between the film and the atmosphere.  In the present section, this question is addressed by  DFT modelling (\textit{ab initio} calculations using the software VASP). It is known that photochromic YHO coatings are obtained experimentally by the partial oxidation of YH\ensuremath{_{2}} films in air. As discussed before, the incorporation of oxygen into YH$_2$ results in the expansion of the YH$_2$ lattice, \emph{i.e.} the lattice parameter (\emph{a}) increases, and hence the XRD peaks corresponding to YHO appear displaced towards lower angles when compared to oxygen-free YH$_2$, Figure \ref{fig3} (a). The oxygen intake also causes the band gap opening: Figure \ref{fig3} (b) shows the experimental transmittance (\emph{T}), reflectance (\emph{R}) and absorbance (\emph{A}) corresponding to YH$_2$ and compared to photochromic YHO. YH$_2$ presents the optical behaviour of a metal but, after the incorporation of oxygen, turns into YHO, a wide-bandgap semiconductor. Taking the crystalline structure of YH$_2$ as the starting point [\emph{fm-3m} and space group symmetry number (SPGN) 225] diverse YHO lattices of stoichiometry YH$_x$O$_y$ have been obtained, Figure \ref{fig3} (c). In particular, multiscale modelling \unskip~\cite{388382:8582469}  predicted the possibility  of  lattices of stoichiometry:
 \textit{(i)} Y\ensuremath{_{4}}H\ensuremath{_{4}}O\ensuremath{_{3}}  111 P\ensuremath{\overline{\text 4}}2\textit{m} and
\textit{(ii)} Y\ensuremath{_{2}}H\ensuremath{_{2}}O 134 P42/\textit{nmm} \unskip~\cite{388382:8582469},
as well as \textit{(iii)}, YHO and SPGN 129 P4/\textit{nmm}, 215 P\ensuremath{\overline{\text 4}}3\textit{m}, 224 P\textit{n}\ensuremath{\overline{\text 3}}\textit{m} and 216 P\ensuremath{\overline{\text 4}}3\textit{m}, amongst others.
Systematic theoretical and experimental studies \unskip~\cite{388382:8582572} pointed to  P\ensuremath{\overline{\text 4}}3\textit{m }with SPGN 216 as the most energetically favourable yttrium oxy-hydride lattice, \textit{i.e.},  stoichiometry $x =1$ and $y = 1$ in YH${_x}$O${_y}$. Please note that YH$_2$ (225), Y$_4$H$_6$O$_2$ (224), Y$_2$H$_2$O (134) as well as Y\ensuremath{_{4}}H\ensuremath{_{4}}O\ensuremath{_{3}} (111), present a metallic character, whereas YHO (216) is predicted to be a wide-band gap semiconductor, as expected experimentally. Besides, the $x =1$ and $y = 1$ stoichiometry is consistent with the results obtained by ion beam analysis \unskip~\cite{388382:8582584}. According to these results, YHO  crystallises into a cubic structure with a lattice constant \textit{a =} 5.29 {\AA}, which corroborates the lattice expansion that takes place in  YH$_2$ (\textit{a =} 5.20 {\AA}) when exposed to air. In particular, the expansion of \emph{a} as well as the opening of the bandgap after air exposure is predicted by DTF, see Figure \ref{fig3} (d), where calculated lattice constant and bandgap are plotted as a function of the Y/O ratio. The predicted value of \emph{a} for YHO is, however, slightly smaller than the experimental value observed (\textit{a =} 5.34 {\AA}) \unskip~\cite{MONGSTAD2011S812, 388382:8582572}; discrepancies may arise due to the thin-film nature of the experimentally studied specimens  as well as to other factors (lattice strain, defects, etc.). In this energetically favourable YHO (216)  lattice, Y, O and H atoms occupy the Wyckoff positions 4c (1/4; 1/4; 1/4), 4a (0, 0, 0) and 4b (1/2, 1/2, 1/2) respectively. Since H and O atoms share tetrahedral sites in the lattice, YHO belong to the emerging family of materials called oxy-hydrides\unskip~\cite{388382:8582572}. The partial oxidation of YH\ensuremath{_{2}}, and hence the formation of YHO, triggers  the expansion of the unit cell volume \unskip~\cite{388382:8582469}. As a consequence of the lattice expansion, the bond distances in YHO will be subjected to oxygen-induced elongation --see Figure \ref{fig3}  (c), where Y-O and Y-H bond lengths, as well as the splitting of the Y 3$d$ states at the conduction band minimum for different  YH$_x$O$_y$ stoichiometries is presented--.

The experimental XPS data and the calculated total density of states of YHO (216) are in good agreement, as shown in Figure \ref{fig3} (c). The opening of a wide band bap as the oxygen atoms are incorporated in the YH\ensuremath{_{2}} structure is also predicted by the \textit{ab initio} calculations, Figure \ref{fig3} (d). However, the model overestimates the band gap, being  4.9 eV the calculated value for YHO (216), approximately 1 eV larger than the experimental  band gap determined in the photochromic films by optical methods \unskip~\cite{388382:8582572,388382:8582486}.  It should be noted that the YHO films, obtained by the oxidation of YH\ensuremath{_{2}} previously prepared by reactive sputtering, are  polycrystalline and multiphase in nature--note the widening of the XRD peaks of YHO when compared to YH$_2$ in Figure \ref{fig3} (a)--. Therefore, the energy band diagram of the material most likerly corresponds to an heterostructure of type-II with staggered band gap. 

The projected DOS for YHO (216) revealed that both O and H atoms strongly contribute to the topmost valence band states. However, they are not hybridized because both H and O atoms are connected to the Y atoms independently from each other. On the other hand, the bottommost conduction band, which in the case of an ideal lattice is three times degenerate, and it is formed mostly by Y \textit{d}-states, in particular t\ensuremath{_{2g}} states, Figure \ref{fig3} (c). This result suggests that the light-induced O release from the film can be caused by the pseudo Jahn-Teller distortion effect: Y atoms are located at the centre of the tetrahedral H and O sublattices and, under illumination, the transfer of electrons from the valence band to the t\ensuremath{_{2g}} bands will turn the YHO (216) lattice unstable \unskip~\cite{388382:8581607}. As the \textit{p} orbitals of O atoms are hybridized with the Y \textit{d} orbitals, the degeneracy of the t\ensuremath{_{2g}} states can be avoided by the removal of oxygen atoms. As a result, an O-deficient unit cell, with smaller lattice constant will be created. As reported in Pishtshev \emph{et al.}\unskip~\cite{388382:8582469} there are many  O-deficient structural arrangements that can be obtained from YHO, being  Y$_4$H$_6$O$_2$ (224) a likely candidate of predicted metallic character and lattice constant $a$ = 5.27 {\AA}  (Figure \ref{fig3}).  In summary, as a result of illumination, metallic domains of smaller lattice constant will be created in the YHO (216) lattice, which result in the in the photohromic effect and the lattice contraction observed experimentally.  The material seems to be able to host the out-diffused O atoms which, in some cases can reach the surface being detected by XPS or even leave the film as demonstrated before. After stopping the illumination, the released O atoms can return to their former positions and the initial optical transparency will be restored.

\section{Conclusions}
 When exposed to air the YH\ensuremath{_2} lattice expands from 5.20 to 5.34 {\AA} due to the incorporation of oxygen. Beyond the lattice expansion, YH\ensuremath{_{2}} turns into YHO which is transparent and photochromic. However, the incorporation of oxygen can be, at least partially, reverted by illumination. According to theoretical calculations, the pseudo Jan-Teller effect may cause the weakening of the Y-O bond by the action of light of adequate wavelength (UV and blue); as a result, some oxygen atoms can be released  leaving behind an oxygen deficient structure responsible for the photochromic darkening of the YHO films. Released oxygen atoms move towards the film surface, causing an enhancement of the hydrophobic properties, being some O atoms able to leave the film. For this reason, the adequate bleaching of the photo-darkened films has to take place in an oxygen rich atmosphere. In summary, YHO ``\textit{breathes}'' as a consequence of illumination: the YHO lattice contraction/expansion under illumination/darkness is accompanied by oxygen release/intake which causes the observed change in the optical properties.

\section*{Acknowledgements}This work has been supported by the Norwegian Research Council through the FRINATEK project 287545, internal project of the Institute for Energy Technology and Turkish Council of Higher Education Board 100/2000 PhD scholarship. The computations have been performed by using the Norwegian Notur supercomputing facilities through the project nn4608k.

\section{BIBLIOGRAPHY}
\bibliography{article.bib}

\begin{thebibliography}{46}%
\makeatletter
\providecommand \@ifxundefined [1]{%
 \@ifx{#1\undefined}
}%
\providecommand \@ifnum [1]{%
 \ifnum #1\expandafter \@firstoftwo
 \else \expandafter \@secondoftwo
 \fi
}%
\providecommand \@ifx [1]{%
 \ifx #1\expandafter \@firstoftwo
 \else \expandafter \@secondoftwo
 \fi
}%
\providecommand \natexlab [1]{#1}%
\providecommand \enquote  [1]{``#1''}%
\providecommand \bibnamefont  [1]{#1}%
\providecommand \bibfnamefont [1]{#1}%
\providecommand \citenamefont [1]{#1}%
\providecommand \href@noop [0]{\@secondoftwo}%
\providecommand \href [0]{\begingroup \@sanitize@url \@href}%
\providecommand \@href[1]{\@@startlink{#1}\@@href}%
\providecommand \@@href[1]{\endgroup#1\@@endlink}%
\providecommand \@sanitize@url [0]{\catcode `\\12\catcode `\$12\catcode
  `\&12\catcode `\#12\catcode `\^12\catcode `\_12\catcode `\%12\relax}%
\providecommand \@@startlink[1]{}%
\providecommand \@@endlink[0]{}%
\providecommand \url  [0]{\begingroup\@sanitize@url \@url }%
\providecommand \@url [1]{\endgroup\@href {#1}{\urlprefix }}%
\providecommand \urlprefix  [0]{URL }%
\providecommand \Eprint [0]{\href }%
\providecommand \doibase [0]{http://dx.doi.org/}%
\providecommand \selectlanguage [0]{\@gobble}%
\providecommand \bibinfo  [0]{\@secondoftwo}%
\providecommand \bibfield  [0]{\@secondoftwo}%
\providecommand \translation [1]{[#1]}%
\providecommand \BibitemOpen [0]{}%
\providecommand \bibitemStop [0]{}%
\providecommand \bibitemNoStop [0]{.\EOS\space}%
\providecommand \EOS [0]{\spacefactor3000\relax}%
\providecommand \BibitemShut  [1]{\csname bibitem#1\endcsname}%
\let\auto@bib@innerbib\@empty
\bibitem [{\citenamefont {Huiberts}\ \emph {et~al.}(1996)\citenamefont
  {Huiberts}, \citenamefont {Griessen}, \citenamefont {Rector}, \citenamefont
  {Wijngaarden}, \citenamefont {Dekker}, \citenamefont {Groot},\ and\
  \citenamefont {Koeman}}]{388382:8582484}%
  \BibitemOpen
  \bibfield  {author} {\bibinfo {author} {\bibfnamefont {J.~N.}\ \bibnamefont
  {Huiberts}}, \bibinfo {author} {\bibfnamefont {R.}~\bibnamefont {Griessen}},
  \bibinfo {author} {\bibfnamefont {J.~H.}\ \bibnamefont {Rector}}, \bibinfo
  {author} {\bibfnamefont {R.~J.}\ \bibnamefont {Wijngaarden}}, \bibinfo
  {author} {\bibfnamefont {J.~P.}\ \bibnamefont {Dekker}}, \bibinfo {author}
  {\bibfnamefont {D.~G.~D.}\ \bibnamefont {Groot}}, \ and\ \bibinfo {author}
  {\bibfnamefont {N.~J.}\ \bibnamefont {Koeman}},\ }\href {10.1038/380231a0}
  {\bibfield  {journal} {\bibinfo  {journal} {{Nature}}\ }\textbf {\bibinfo
  {volume} {380}},\ \bibinfo {pages} {231} (\bibinfo {year}
  {1996})}\BibitemShut {NoStop}%
\bibitem [{\citenamefont {Nafezarefi}\ \emph {et~al.}(2017)\citenamefont
  {Nafezarefi}, \citenamefont {Schreuders}, \citenamefont {Dam},\ and\
  \citenamefont {Cornelius}}]{388382:8582534}%
  \BibitemOpen
  \bibfield  {author} {\bibinfo {author} {\bibfnamefont {F.}~\bibnamefont
  {Nafezarefi}}, \bibinfo {author} {\bibfnamefont {H.}~\bibnamefont
  {Schreuders}}, \bibinfo {author} {\bibfnamefont {B.}~\bibnamefont {Dam}}, \
  and\ \bibinfo {author} {\bibfnamefont {S.}~\bibnamefont {Cornelius}},\ }\href
  {10.1063/1.4995081} {\bibfield  {journal} {\bibinfo  {journal} {{Applied
  Physics Letters}}\ }\textbf {\bibinfo {volume} {111}},\ \bibinfo {pages} {3}
  (\bibinfo {year} {2017})}\BibitemShut {NoStop}%
\bibitem [{\citenamefont {Montero}\ \emph {et~al.}(2018)\citenamefont
  {Montero}, \citenamefont {Martinsen}, \citenamefont {Lelis}, \citenamefont
  {Karazhanov}, \citenamefont {Hauback},\ and\ \citenamefont
  {Marstein}}]{388382:8582572}%
  \BibitemOpen
  \bibfield  {author} {\bibinfo {author} {\bibfnamefont {J.}~\bibnamefont
  {Montero}}, \bibinfo {author} {\bibfnamefont {F.~A.}\ \bibnamefont
  {Martinsen}}, \bibinfo {author} {\bibfnamefont {M.}~\bibnamefont {Lelis}},
  \bibinfo {author} {\bibfnamefont {S.~Z.}\ \bibnamefont {Karazhanov}},
  \bibinfo {author} {\bibfnamefont {B.~C.}\ \bibnamefont {Hauback}}, \ and\
  \bibinfo {author} {\bibfnamefont {E.~S.}\ \bibnamefont {Marstein}},\ }\href
  {10.1016/j.solmat.2017.02.001} {\bibfield  {journal} {\bibinfo  {journal}
  {{Solar Energy Materials and Solar Cells}}\ }\textbf {\bibinfo {volume}
  {177}},\ \bibinfo {pages} {106} (\bibinfo {year} {2018})}\BibitemShut
  {NoStop}%
\bibitem [{\citenamefont {La}\ \emph {et~al.}(2018)\citenamefont {La},
  \citenamefont {Li}, \citenamefont {Sha}, \citenamefont {Bao},\ and\
  \citenamefont {Jin}}]{388382:8582483}%
  \BibitemOpen
  \bibfield  {author} {\bibinfo {author} {\bibfnamefont {M.}~\bibnamefont
  {La}}, \bibinfo {author} {\bibfnamefont {N.}~\bibnamefont {Li}}, \bibinfo
  {author} {\bibfnamefont {R.}~\bibnamefont {Sha}}, \bibinfo {author}
  {\bibfnamefont {S.}~\bibnamefont {Bao}}, \ and\ \bibinfo {author}
  {\bibfnamefont {P.}~\bibnamefont {Jin}},\ }\href
  {10.1016/j.scriptamat.2017.08.020} {\bibfield  {journal} {\bibinfo  {journal}
  {{Scripta Materialia}}\ }\textbf {\bibinfo {volume} {142}},\ \bibinfo {pages}
  {36} (\bibinfo {year} {2018})}\BibitemShut {NoStop}%
\bibitem [{\citenamefont {Miniotas}\ \emph {et~al.}(2000)\citenamefont
  {Miniotas}, \citenamefont {Hj{\"{o}}rvarsson}, \citenamefont {Douysset},\
  and\ \citenamefont {Nostell}}]{388382:8582482}%
  \BibitemOpen
  \bibfield  {author} {\bibinfo {author} {\bibfnamefont {A.}~\bibnamefont
  {Miniotas}}, \bibinfo {author} {\bibfnamefont {B.}~\bibnamefont
  {Hj{\"{o}}rvarsson}}, \bibinfo {author} {\bibfnamefont {L.}~\bibnamefont
  {Douysset}}, \ and\ \bibinfo {author} {\bibfnamefont {P.}~\bibnamefont
  {Nostell}},\ }\href {10.1063/1.126253} {\bibfield  {journal} {\bibinfo
  {journal} {{Applied Physics Letters}}\ }\textbf {\bibinfo {volume} {76}},\
  \bibinfo {pages} {2056} (\bibinfo {year} {2000})}\BibitemShut {NoStop}%
\bibitem [{\citenamefont {Mongstad}\ \emph
  {et~al.}(2011{\natexlab{a}})\citenamefont {Mongstad}, \citenamefont
  {Platzer-Bj{\"{o}}rkman}, \citenamefont {Maehlen}, \citenamefont {Mooij},
  \citenamefont {Pivak}, \citenamefont {Dam}, \citenamefont {Marstein},
  \citenamefont {Hauback},\ and\ \citenamefont {Karazhanov}}]{388382:8582563}%
  \BibitemOpen
  \bibfield  {author} {\bibinfo {author} {\bibfnamefont {T.}~\bibnamefont
  {Mongstad}}, \bibinfo {author} {\bibfnamefont {C.}~\bibnamefont
  {Platzer-Bj{\"{o}}rkman}}, \bibinfo {author} {\bibfnamefont {J.~P.}\
  \bibnamefont {Maehlen}}, \bibinfo {author} {\bibfnamefont {L.~P.~A.}\
  \bibnamefont {Mooij}}, \bibinfo {author} {\bibfnamefont {Y.}~\bibnamefont
  {Pivak}}, \bibinfo {author} {\bibfnamefont {B.}~\bibnamefont {Dam}}, \bibinfo
  {author} {\bibfnamefont {E.~S.}\ \bibnamefont {Marstein}}, \bibinfo {author}
  {\bibfnamefont {B.~C.}\ \bibnamefont {Hauback}}, \ and\ \bibinfo {author}
  {\bibfnamefont {S.~Z.}\ \bibnamefont {Karazhanov}},\ }\href
  {10.1016/j.solmat.2011.08.018} {\bibfield  {journal} {\bibinfo  {journal}
  {{Solar Energy Materials and Solar Cells}}\ }\textbf {\bibinfo {volume}
  {95}},\ \bibinfo {pages} {3596} (\bibinfo {year}
  {2011}{\natexlab{a}})}\BibitemShut {NoStop}%
\bibitem [{\citenamefont {Ohmura}\ \emph {et~al.}(2007)\citenamefont {Ohmura},
  \citenamefont {MacHida}, \citenamefont {Watanuki}, \citenamefont {Aoki},
  \citenamefont {Nakano},\ and\ \citenamefont {Takemura}}]{388382:8582546}%
  \BibitemOpen
  \bibfield  {author} {\bibinfo {author} {\bibfnamefont {A.}~\bibnamefont
  {Ohmura}}, \bibinfo {author} {\bibfnamefont {A.}~\bibnamefont {MacHida}},
  \bibinfo {author} {\bibfnamefont {T.}~\bibnamefont {Watanuki}}, \bibinfo
  {author} {\bibfnamefont {K.}~\bibnamefont {Aoki}}, \bibinfo {author}
  {\bibfnamefont {S.}~\bibnamefont {Nakano}}, \ and\ \bibinfo {author}
  {\bibfnamefont {K.}~\bibnamefont {Takemura}},\ }\href {10.1063/1.2794755}
  {\bibfield  {journal} {\bibinfo  {journal} {{Applied Physics Letters}}\
  }\textbf {\bibinfo {volume} {91}} (\bibinfo {year} {2007})}\BibitemShut
  {NoStop}%
\bibitem [{\citenamefont {Towns}(2016)}]{388382:8581903}%
  \BibitemOpen
  \bibfield  {author} {\bibinfo {author} {\bibfnamefont {A.}~\bibnamefont
  {Towns}},\ }\href@noop {} {\emph {\bibinfo {title} {{Applied Photochemistry.
  In: Applied Photochemistry: When Light Meets Molecules}}}}\ (\bibinfo
  {publisher} {Springer International Publishing},\ \bibinfo {address}
  {Switzerland},\ \bibinfo {year} {2016})\ pp.\ \bibinfo {pages}
  {227--79}\BibitemShut {NoStop}%
\bibitem [{\citenamefont {Montero}\ \emph {et~al.}(2017)\citenamefont
  {Montero}, \citenamefont {Martinsen}, \citenamefont {Garc\'{\i}a-Tecedor},
  \citenamefont {Karazhanov}, \citenamefont {Maestre}, \citenamefont
  {Hauback},\ and\ \citenamefont {Marstein}}]{388382:8582573}%
  \BibitemOpen
  \bibfield  {author} {\bibinfo {author} {\bibfnamefont {J.}~\bibnamefont
  {Montero}}, \bibinfo {author} {\bibfnamefont {F.~A.}\ \bibnamefont
  {Martinsen}}, \bibinfo {author} {\bibfnamefont {M.}~\bibnamefont
  {Garc\'{\i}a-Tecedor}}, \bibinfo {author} {\bibfnamefont {S.~Z.}\
  \bibnamefont {Karazhanov}}, \bibinfo {author} {\bibfnamefont
  {D.}~\bibnamefont {Maestre}}, \bibinfo {author} {\bibfnamefont
  {B.}~\bibnamefont {Hauback}}, \ and\ \bibinfo {author} {\bibfnamefont
  {E.~S.}\ \bibnamefont {Marstein}},\ }\href {10.1103/PhysRevB.95.201301}
  {\bibfield  {journal} {\bibinfo  {journal} {{Physical Review B}}\ }\textbf
  {\bibinfo {volume} {95}},\ \bibinfo {pages} {1} (\bibinfo {year}
  {2017})}\BibitemShut {NoStop}%
\bibitem [{\citenamefont {Zenkin}\ \emph {et~al.}(2014)\citenamefont {Zenkin},
  \citenamefont {Kos},\ and\ \citenamefont {Musil}}]{388382:8582513}%
  \BibitemOpen
  \bibfield  {author} {\bibinfo {author} {\bibfnamefont {S.}~\bibnamefont
  {Zenkin}}, \bibinfo {author} {\bibfnamefont {S.}~\bibnamefont {Kos}}, \ and\
  \bibinfo {author} {\bibfnamefont {J.}~\bibnamefont {Musil}},\ }\href
  {10.1111/jace.13165} {\bibfield  {journal} {\bibinfo  {journal} {{Journal of
  the American Ceramic Society}}\ }\textbf {\bibinfo {volume} {97}},\ \bibinfo
  {pages} {2713} (\bibinfo {year} {2014})}\BibitemShut {NoStop}%
\bibitem [{\citenamefont {Azimi}\ \emph {et~al.}(2013)\citenamefont {Azimi},
  \citenamefont {Dhiman}, \citenamefont {Kwon}, \citenamefont {Paxson},\ and\
  \citenamefont {Varanasi}}]{388382:8582588}%
  \BibitemOpen
  \bibfield  {author} {\bibinfo {author} {\bibfnamefont {G.}~\bibnamefont
  {Azimi}}, \bibinfo {author} {\bibfnamefont {R.}~\bibnamefont {Dhiman}},
  \bibinfo {author} {\bibfnamefont {H.~M.}\ \bibnamefont {Kwon}}, \bibinfo
  {author} {\bibfnamefont {A.~T.}\ \bibnamefont {Paxson}}, \ and\ \bibinfo
  {author} {\bibfnamefont {K.~K.}\ \bibnamefont {Varanasi}},\ }\href
  {10.1038/nmat3545} {\bibfield  {journal} {\bibinfo  {journal} {{Nature
  Materials}}\ }\textbf {\bibinfo {volume} {12}},\ \bibinfo {pages} {315}
  (\bibinfo {year} {2013})}\BibitemShut {NoStop}%
\bibitem [{\citenamefont {Ho}\ \emph {et~al.}(2007)\citenamefont {Ho},
  \citenamefont {Kwak}, \citenamefont {Dong}, \citenamefont {Seung},\ and\
  \citenamefont {Cho}}]{388382:8582538}%
  \BibitemOpen
  \bibfield  {author} {\bibinfo {author} {\bibfnamefont {S.~L.}\ \bibnamefont
  {Ho}}, \bibinfo {author} {\bibfnamefont {D.}~\bibnamefont {Kwak}}, \bibinfo
  {author} {\bibfnamefont {Y.~L.}\ \bibnamefont {Dong}}, \bibinfo {author}
  {\bibfnamefont {G.~L.}\ \bibnamefont {Seung}}, \ and\ \bibinfo {author}
  {\bibfnamefont {K.}~\bibnamefont {Cho}},\ }\href {10.1021/ja0692579}
  {\bibfield  {journal} {\bibinfo  {journal} {{Journal of the American Chemical
  Society}}\ }\textbf {\bibinfo {volume} {129}},\ \bibinfo {pages} {4128}
  (\bibinfo {year} {2007})}\BibitemShut {NoStop}%
\bibitem [{\citenamefont {Feng}\ \emph {et~al.}(2004)\citenamefont {Feng},
  \citenamefont {Feng}, \citenamefont {Jin}, \citenamefont {Zhai},
  \citenamefont {Jiang},\ and\ \citenamefont {Zhu}}]{388382:8582580}%
  \BibitemOpen
  \bibfield  {author} {\bibinfo {author} {\bibfnamefont {X.}~\bibnamefont
  {Feng}}, \bibinfo {author} {\bibfnamefont {L.}~\bibnamefont {Feng}}, \bibinfo
  {author} {\bibfnamefont {M.}~\bibnamefont {Jin}}, \bibinfo {author}
  {\bibfnamefont {J.}~\bibnamefont {Zhai}}, \bibinfo {author} {\bibfnamefont
  {L.}~\bibnamefont {Jiang}}, \ and\ \bibinfo {author} {\bibfnamefont
  {D.}~\bibnamefont {Zhu}},\ }\href {10.1021/ja038636o} {\bibfield  {journal}
  {\bibinfo  {journal} {{Journal of the American Chemical Society}}\ }\textbf
  {\bibinfo {volume} {126}},\ \bibinfo {pages} {62} (\bibinfo {year}
  {2004})}\BibitemShut {NoStop}%
\bibitem [{\citenamefont {Yadav}\ \emph {et~al.}(2016)\citenamefont {Yadav},
  \citenamefont {Mehta}, \citenamefont {Bhattacharya},\ and\ \citenamefont
  {Singh}}]{388382:8582537}%
  \BibitemOpen
  \bibfield  {author} {\bibinfo {author} {\bibfnamefont {K.}~\bibnamefont
  {Yadav}}, \bibinfo {author} {\bibfnamefont {B.~R.}\ \bibnamefont {Mehta}},
  \bibinfo {author} {\bibfnamefont {S.}~\bibnamefont {Bhattacharya}}, \ and\
  \bibinfo {author} {\bibfnamefont {J.~P.}\ \bibnamefont {Singh}},\ }\href
  {10.1038/srep35073} {\bibfield  {journal} {\bibinfo  {journal} {{Scientific
  Reports}}\ }\textbf {\bibinfo {volume} {6}} (\bibinfo {year}
  {2016})}\BibitemShut {NoStop}%
\bibitem [{\citenamefont {Montero}\ and\ \citenamefont
  {Karazhanov}(2018)}]{388382:8582486}%
  \BibitemOpen
  \bibfield  {author} {\bibinfo {author} {\bibfnamefont {J.}~\bibnamefont
  {Montero}}\ and\ \bibinfo {author} {\bibfnamefont {S.~Z.}\ \bibnamefont
  {Karazhanov}},\ }\href {10.1002/pssa.201701039} {\bibfield  {journal}
  {\bibinfo  {journal} {{Physica Status Solidi (A) Applications and Materials
  Science}}\ }\textbf {\bibinfo {volume} {215}},\ \bibinfo {pages} {1}
  (\bibinfo {year} {2018})}\BibitemShut {NoStop}%
\bibitem [{\citenamefont {van Oss}\ \emph {et~al.}(1988)\citenamefont {van
  Oss}, \citenamefont {Chaudhury},\ and\ \citenamefont
  {Good}}]{388382:8582545}%
  \BibitemOpen
  \bibfield  {author} {\bibinfo {author} {\bibfnamefont {C.~J.}\ \bibnamefont
  {van Oss}}, \bibinfo {author} {\bibfnamefont {M.~K.}\ \bibnamefont
  {Chaudhury}}, \ and\ \bibinfo {author} {\bibfnamefont {R.~J.}\ \bibnamefont
  {Good}},\ }\href {10.1021/cr00088a006} {\bibfield  {journal} {\bibinfo
  {journal} {{Chemical Reviews}}\ }\textbf {\bibinfo {volume} {88}},\ \bibinfo
  {pages} {927} (\bibinfo {year} {1988})}\BibitemShut {NoStop}%
\bibitem [{\citenamefont {van Oss}(1993)}]{388382:8582531}%
  \BibitemOpen
  \bibfield  {author} {\bibinfo {author} {\bibfnamefont {C.~J.}\ \bibnamefont
  {van Oss}},\ }\href {10.1016/0927-7757(93)80308-2} {\bibfield  {journal}
  {\bibinfo  {journal} {{Colloids and Surfaces A: Physicochemical and
  Engineering Aspects}}\ } (\bibinfo {year} {1993})}\BibitemShut {NoStop}%
\bibitem [{\citenamefont {G}\ and\ \citenamefont
  {J}(1996{\natexlab{a}})}]{388382:8581928}%
  \BibitemOpen
  \bibfield  {author} {\bibinfo {author} {\bibfnamefont {K.}~\bibnamefont {G}}\
  and\ \bibinfo {author} {\bibfnamefont {F.}~\bibnamefont {J}},\ }\href
  {10.1103/PhysRevB.54.11169} {\bibfield  {journal} {\bibinfo  {journal} {{Phys
  Rev B - Condens Matter Mater Phys}}\ }\textbf {\bibinfo {volume} {54}},\
  \bibinfo {pages} {11169} (\bibinfo {year} {1996}{\natexlab{a}})}\BibitemShut
  {NoStop}%
\bibitem [{\citenamefont {G}\ and\ \citenamefont
  {J}(1996{\natexlab{b}})}]{388382:8581929}%
  \BibitemOpen
  \bibfield  {author} {\bibinfo {author} {\bibfnamefont {K.}~\bibnamefont {G}}\
  and\ \bibinfo {author} {\bibfnamefont {F.}~\bibnamefont {J}},\ }\href
  {10.1016/0927-0256(96)00008-0} {\bibfield  {journal} {\bibinfo  {journal}
  {{Comput Mater Sci}}\ }\textbf {\bibinfo {volume} {6}},\ \bibinfo {pages}
  {15} (\bibinfo {year} {1996}{\natexlab{b}})}\BibitemShut {NoStop}%
\bibitem [{\citenamefont {G}\ and\ \citenamefont {J}(1994)}]{388382:8581930}%
  \BibitemOpen
  \bibfield  {author} {\bibinfo {author} {\bibfnamefont {K.}~\bibnamefont {G}}\
  and\ \bibinfo {author} {\bibfnamefont {H.}~\bibnamefont {J}},\ }\href
  {10.1103/PhysRevB.49.14251} {\bibfield  {journal} {\bibinfo  {journal} {{Phys
  Rev B}}\ }\textbf {\bibinfo {volume} {49}},\ \bibinfo {pages} {14251}
  (\bibinfo {year} {1994})}\BibitemShut {NoStop}%
\bibitem [{\citenamefont {Bl{\"{o}}chl}(1994)}]{388382:8582478}%
  \BibitemOpen
  \bibfield  {author} {\bibinfo {author} {\bibfnamefont {P.~E.}\ \bibnamefont
  {Bl{\"{o}}chl}},\ }\href {10.1103/PhysRevB.50.17953} {\bibfield  {journal}
  {\bibinfo  {journal} {{Physical Review B}}\ }\textbf {\bibinfo {volume}
  {50}},\ \bibinfo {pages} {17953} (\bibinfo {year} {1994})}\BibitemShut
  {NoStop}%
\bibitem [{\citenamefont {Kresse}\ and\ \citenamefont
  {Joubert}(1999)}]{388382:8582477}%
  \BibitemOpen
  \bibfield  {author} {\bibinfo {author} {\bibfnamefont {G.}~\bibnamefont
  {Kresse}}\ and\ \bibinfo {author} {\bibfnamefont {D.}~\bibnamefont
  {Joubert}},\ }\href {10.1103/PhysRevB.59.1758} {\bibfield  {journal}
  {\bibinfo  {journal} {{Physical Review B}}\ }\textbf {\bibinfo {volume}
  {59}},\ \bibinfo {pages} {1758} (\bibinfo {year} {1999})}\BibitemShut
  {NoStop}%
\bibitem [{\citenamefont {G}\ and\ \citenamefont {J}(1993)}]{388382:8581933}%
  \BibitemOpen
  \bibfield  {author} {\bibinfo {author} {\bibfnamefont {K.}~\bibnamefont {G}}\
  and\ \bibinfo {author} {\bibfnamefont {H.}~\bibnamefont {J}}\ }(\bibinfo
  {year} {1993})\ pp.\ \bibinfo {pages} {558--561}\BibitemShut {NoStop}%
\bibitem [{\citenamefont {Perdew}\ \emph {et~al.}(1996)\citenamefont {Perdew},
  \citenamefont {Burke},\ and\ \citenamefont {Ernzerhof}}]{388382:8582475}%
  \BibitemOpen
  \bibfield  {author} {\bibinfo {author} {\bibfnamefont {J.~P.}\ \bibnamefont
  {Perdew}}, \bibinfo {author} {\bibfnamefont {K.}~\bibnamefont {Burke}}, \
  and\ \bibinfo {author} {\bibfnamefont {M.}~\bibnamefont {Ernzerhof}},\ }\href
  {10.1103/PhysRevLett.77.3865} {\bibfield  {journal} {\bibinfo  {journal}
  {{Physical Review Letters}}\ }\textbf {\bibinfo {volume} {77}},\ \bibinfo
  {pages} {3865} (\bibinfo {year} {1996})}\BibitemShut {NoStop}%
\bibitem [{\citenamefont {Heyd}\ \emph {et~al.}(2003)\citenamefont {Heyd},
  \citenamefont {Scuseria},\ and\ \citenamefont {Ernzerhof}}]{388382:8581935}%
  \BibitemOpen
  \bibfield  {author} {\bibinfo {author} {\bibfnamefont {J.}~\bibnamefont
  {Heyd}}, \bibinfo {author} {\bibfnamefont {G.~E.}\ \bibnamefont {Scuseria}},
  \ and\ \bibinfo {author} {\bibfnamefont {M.}~\bibnamefont {Ernzerhof}},\
  }\href {10.1063/1.1564060} {\bibfield  {journal} {\bibinfo  {journal} {{J
  Chem Phys}}\ }\textbf {\bibinfo {volume} {118}},\ \bibinfo {pages} {8207}
  (\bibinfo {year} {2003})}\BibitemShut {NoStop}%
\bibitem [{\citenamefont {Krukau}\ \emph {et~al.}(2006)\citenamefont {Krukau},
  \citenamefont {Vydrov}, \citenamefont {Izmaylov},\ and\ \citenamefont
  {Scuseria}}]{388382:8582473}%
  \BibitemOpen
  \bibfield  {author} {\bibinfo {author} {\bibfnamefont {A.~V.}\ \bibnamefont
  {Krukau}}, \bibinfo {author} {\bibfnamefont {O.~A.}\ \bibnamefont {Vydrov}},
  \bibinfo {author} {\bibfnamefont {A.~F.}\ \bibnamefont {Izmaylov}}, \ and\
  \bibinfo {author} {\bibfnamefont {G.~E.}\ \bibnamefont {Scuseria}},\ }\href
  {10.1063/1.2404663} {\bibfield  {journal} {\bibinfo  {journal} {{The Journal
  of Chemical Physics}}\ }\textbf {\bibinfo {volume} {125}},\ \bibinfo {pages}
  {224106} (\bibinfo {year} {2006})}\BibitemShut {NoStop}%
\bibitem [{\citenamefont {Henderson}\ \emph {et~al.}(2011)\citenamefont
  {Henderson}, \citenamefont {Paier},\ and\ \citenamefont
  {Scuseria}}]{388382:8582472}%
  \BibitemOpen
  \bibfield  {author} {\bibinfo {author} {\bibfnamefont {T.~M.}\ \bibnamefont
  {Henderson}}, \bibinfo {author} {\bibfnamefont {J.}~\bibnamefont {Paier}}, \
  and\ \bibinfo {author} {\bibfnamefont {G.~E.}\ \bibnamefont {Scuseria}},\
  }\href {10.1002/pssb.201046303} {\bibfield  {journal} {\bibinfo  {journal}
  {{physica status solidi (b)}}\ }\textbf {\bibinfo {volume} {248}},\ \bibinfo
  {pages} {767} (\bibinfo {year} {2011})}\BibitemShut {NoStop}%
\bibitem [{\citenamefont {Li}\ \emph {et~al.}(2010)\citenamefont {Li},
  \citenamefont {Niklasson},\ and\ \citenamefont {Granqvist}}]{388382:8582468}%
  \BibitemOpen
  \bibfield  {author} {\bibinfo {author} {\bibfnamefont {S.~Y.}\ \bibnamefont
  {Li}}, \bibinfo {author} {\bibfnamefont {G.~A.}\ \bibnamefont {Niklasson}}, \
  and\ \bibinfo {author} {\bibfnamefont {C.~G.}\ \bibnamefont {Granqvist}},\
  }\href {10.1063/1.3487980} {\bibfield  {journal} {\bibinfo  {journal}
  {{Journal of Applied Physics}}\ }\textbf {\bibinfo {volume} {108}} (\bibinfo
  {year} {2010})}\BibitemShut {NoStop}%
\bibitem [{\citenamefont {Imanaka}(2004)}]{388382:8582533}%
  \BibitemOpen
  \bibfield  {author} {\bibinfo {author} {\bibfnamefont {N.}~\bibnamefont
  {Imanaka}},\ }\href@noop {} {\bibfield  {journal} {\bibinfo  {journal}
  {{Binary Rare Earth Oxides}}\ ,\ \bibinfo {pages} {111}} (\bibinfo {year}
  {2004})}\BibitemShut {NoStop}%
\bibitem [{\citenamefont {Moldarev}\ \emph {et~al.}(2018)\citenamefont
  {Moldarev}, \citenamefont {Primetzhofer}, \citenamefont {You}, \citenamefont
  {Karazhanov}, \citenamefont {Montero}, \citenamefont {Martinsen},
  \citenamefont {Mongstad}, \citenamefont {Marstein},\ and\ \citenamefont
  {Wolff}}]{388382:8582584}%
  \BibitemOpen
  \bibfield  {author} {\bibinfo {author} {\bibfnamefont {D.}~\bibnamefont
  {Moldarev}}, \bibinfo {author} {\bibfnamefont {D.}~\bibnamefont
  {Primetzhofer}}, \bibinfo {author} {\bibfnamefont {C.~C.}\ \bibnamefont
  {You}}, \bibinfo {author} {\bibfnamefont {S.~Z.}\ \bibnamefont {Karazhanov}},
  \bibinfo {author} {\bibfnamefont {J.}~\bibnamefont {Montero}}, \bibinfo
  {author} {\bibfnamefont {F.}~\bibnamefont {Martinsen}}, \bibinfo {author}
  {\bibfnamefont {T.}~\bibnamefont {Mongstad}}, \bibinfo {author}
  {\bibfnamefont {E.~S.}\ \bibnamefont {Marstein}}, \ and\ \bibinfo {author}
  {\bibfnamefont {M.}~\bibnamefont {Wolff}},\ }\href
  {10.1016/j.solmat.2017.05.052} {\bibfield  {journal} {\bibinfo  {journal}
  {{Solar Energy Materials and Solar Cells}}\ }\textbf {\bibinfo {volume}
  {177}},\ \bibinfo {pages} {66} (\bibinfo {year} {2018})}\BibitemShut
  {NoStop}%
\bibitem [{\citenamefont {Feng}\ \emph {et~al.}(2005)\citenamefont {Feng},
  \citenamefont {Zhai},\ and\ \citenamefont {Jiang}}]{388382:8582536}%
  \BibitemOpen
  \bibfield  {author} {\bibinfo {author} {\bibfnamefont {X.}~\bibnamefont
  {Feng}}, \bibinfo {author} {\bibfnamefont {J.}~\bibnamefont {Zhai}}, \ and\
  \bibinfo {author} {\bibfnamefont {L.}~\bibnamefont {Jiang}},\ }\href
  {10.1002/anie.200501337} {\bibfield  {journal} {\bibinfo  {journal}
  {{Angewandte Chemie - International Edition}}\ }\textbf {\bibinfo {volume}
  {44}},\ \bibinfo {pages} {5115} (\bibinfo {year} {2005})}\BibitemShut
  {NoStop}%
\bibitem [{\citenamefont {Wang}\ \emph {et~al.}(2006)\citenamefont {Wang},
  \citenamefont {Feng}, \citenamefont {Yao},\ and\ \citenamefont
  {Jiang}}]{388382:8582579}%
  \BibitemOpen
  \bibfield  {author} {\bibinfo {author} {\bibfnamefont {S.}~\bibnamefont
  {Wang}}, \bibinfo {author} {\bibfnamefont {X.}~\bibnamefont {Feng}}, \bibinfo
  {author} {\bibfnamefont {J.}~\bibnamefont {Yao}}, \ and\ \bibinfo {author}
  {\bibfnamefont {L.}~\bibnamefont {Jiang}},\ }\href {10.1002/anie.200502061}
  {\bibfield  {journal} {\bibinfo  {journal} {{Angewandte Chemie -
  International Edition}}\ }\textbf {\bibinfo {volume} {45}},\ \bibinfo {pages}
  {1264} (\bibinfo {year} {2006})}\BibitemShut {NoStop}%
\bibitem [{\citenamefont {Zhu}\ \emph {et~al.}(2006)\citenamefont {Zhu},
  \citenamefont {Feng}, \citenamefont {Feng},\ and\ \citenamefont
  {Jiang}}]{388382:8582578}%
  \BibitemOpen
  \bibfield  {author} {\bibinfo {author} {\bibfnamefont {W.}~\bibnamefont
  {Zhu}}, \bibinfo {author} {\bibfnamefont {X.}~\bibnamefont {Feng}}, \bibinfo
  {author} {\bibfnamefont {L.}~\bibnamefont {Feng}}, \ and\ \bibinfo {author}
  {\bibfnamefont {L.}~\bibnamefont {Jiang}},\ }\href {10.1039/b603634a}
  {\bibfield  {journal} {\bibinfo  {journal} {{Chemical Communications}}\ ,\
  \bibinfo {pages} {2753}} (\bibinfo {year} {2006})}\BibitemShut {NoStop}%
\bibitem [{\citenamefont {de~Sun}\ \emph {et~al.}(2001)\citenamefont {de~Sun},
  \citenamefont {Nakajima}, \citenamefont {Fujishima}, \citenamefont
  {Watanabe},\ and\ \citenamefont {Hashimoto}}]{388382:8582518}%
  \BibitemOpen
  \bibfield  {author} {\bibinfo {author} {\bibfnamefont {R.}~\bibnamefont
  {de~Sun}}, \bibinfo {author} {\bibfnamefont {A.}~\bibnamefont {Nakajima}},
  \bibinfo {author} {\bibfnamefont {A.}~\bibnamefont {Fujishima}}, \bibinfo
  {author} {\bibfnamefont {T.}~\bibnamefont {Watanabe}}, \ and\ \bibinfo
  {author} {\bibfnamefont {K.}~\bibnamefont {Hashimoto}},\ }\href
  {10.1021/jp002525j} {\bibfield  {journal} {\bibinfo  {journal} {{The Journal
  of Physical Chemistry B}}\ }\textbf {\bibinfo {volume} {105}},\ \bibinfo
  {pages} {1984} (\bibinfo {year} {2001})}\BibitemShut {NoStop}%
\bibitem [{\citenamefont {Kung}(1989)}]{388382:8581940}%
  \BibitemOpen
  \bibfield  {author} {\bibinfo {author} {\bibfnamefont {H.~H.}\ \bibnamefont
  {Kung}},\ }in\ \href@noop {} {\emph {\bibinfo {booktitle} {Transition Metal
  Oxides}}},\ \bibinfo {series} {Studies in Surface Science and Catalysis},
  Vol.~\bibinfo {volume} {45}\ (\bibinfo  {publisher} {Elsevier, The
  Netherlands},\ \bibinfo {year} {1989})\ pp.\ \bibinfo {pages} {53 --
  71}\BibitemShut {NoStop}%
\bibitem [{\citenamefont {Khan}\ \emph {et~al.}(2015)\citenamefont {Khan},
  \citenamefont {Azimi}, \citenamefont {Yildiz},\ and\ \citenamefont
  {Varanasi}}]{388382:8582577}%
  \BibitemOpen
  \bibfield  {author} {\bibinfo {author} {\bibfnamefont {S.}~\bibnamefont
  {Khan}}, \bibinfo {author} {\bibfnamefont {G.}~\bibnamefont {Azimi}},
  \bibinfo {author} {\bibfnamefont {B.}~\bibnamefont {Yildiz}}, \ and\ \bibinfo
  {author} {\bibfnamefont {K.~K.}\ \bibnamefont {Varanasi}},\ }\href
  {10.1063/1.4907756} {\bibfield  {journal} {\bibinfo  {journal} {{Applied
  Physics Letters}}\ }\textbf {\bibinfo {volume} {106}},\ \bibinfo {pages}
  {061601} (\bibinfo {year} {2015})}\BibitemShut {NoStop}%
\bibitem [{\citenamefont {Beamson}\ and\ \citenamefont
  {Briggs}(1993)}]{Beamson93}%
  \BibitemOpen
  \bibfield  {author} {\bibinfo {author} {\bibfnamefont {G.}~\bibnamefont
  {Beamson}}\ and\ \bibinfo {author} {\bibfnamefont {D.}~\bibnamefont
  {Briggs}},\ }\href@noop {} {\bibfield  {journal} {\bibinfo  {journal}
  {Journal of Chemical Education}\ }\textbf {\bibinfo {volume} {70}},\ \bibinfo
  {pages} {A25} (\bibinfo {year} {1993})}\BibitemShut {NoStop}%
\bibitem [{\citenamefont {Preston}\ \emph {et~al.}(2014)\citenamefont
  {Preston}, \citenamefont {Miljkovic}, \citenamefont {Sack}, \citenamefont
  {Enright}, \citenamefont {Queeney},\ and\ \citenamefont
  {Wang}}]{388382:8582495}%
  \BibitemOpen
  \bibfield  {author} {\bibinfo {author} {\bibfnamefont {D.~J.}\ \bibnamefont
  {Preston}}, \bibinfo {author} {\bibfnamefont {N.}~\bibnamefont {Miljkovic}},
  \bibinfo {author} {\bibfnamefont {J.}~\bibnamefont {Sack}}, \bibinfo {author}
  {\bibfnamefont {R.}~\bibnamefont {Enright}}, \bibinfo {author} {\bibfnamefont
  {J.}~\bibnamefont {Queeney}}, \ and\ \bibinfo {author} {\bibfnamefont
  {E.~N.}\ \bibnamefont {Wang}},\ }\href {10.1063/1.4886410} {\bibfield
  {journal} {\bibinfo  {journal} {{Applied Physics Letters}}\ }\textbf
  {\bibinfo {volume} {105}},\ \bibinfo {pages} {011601} (\bibinfo {year}
  {2014})}\BibitemShut {NoStop}%
\bibitem [{\citenamefont {Lundy}\ \emph {et~al.}(2017)\citenamefont {Lundy},
  \citenamefont {Byrne}, \citenamefont {Bogan}, \citenamefont {Nolan},
  \citenamefont {Collins}, \citenamefont {Dalton},\ and\ \citenamefont
  {Enright}}]{388382:8582512}%
  \BibitemOpen
  \bibfield  {author} {\bibinfo {author} {\bibfnamefont {R.}~\bibnamefont
  {Lundy}}, \bibinfo {author} {\bibfnamefont {C.}~\bibnamefont {Byrne}},
  \bibinfo {author} {\bibfnamefont {J.}~\bibnamefont {Bogan}}, \bibinfo
  {author} {\bibfnamefont {K.}~\bibnamefont {Nolan}}, \bibinfo {author}
  {\bibfnamefont {M.~N.}\ \bibnamefont {Collins}}, \bibinfo {author}
  {\bibfnamefont {E.}~\bibnamefont {Dalton}}, \ and\ \bibinfo {author}
  {\bibfnamefont {R.}~\bibnamefont {Enright}},\ }\href {10.1021/acsami.7b01515}
  {\bibfield  {journal} {\bibinfo  {journal} {{ACS Applied Materials and
  Interfaces}}\ }\textbf {\bibinfo {volume} {9}},\ \bibinfo {pages} {13751}
  (\bibinfo {year} {2017})}\BibitemShut {NoStop}%
\bibitem [{\citenamefont {K{\"{u}}lah}\ \emph {et~al.}(2017)\citenamefont
  {K{\"{u}}lah}, \citenamefont {Marot}, \citenamefont {Steiner}, \citenamefont
  {Romanyuk}, \citenamefont {Jung}, \citenamefont {W{\"{a}}ckerlin},\ and\
  \citenamefont {Meyer}}]{388382:8582551}%
  \BibitemOpen
  \bibfield  {author} {\bibinfo {author} {\bibfnamefont {E.}~\bibnamefont
  {K{\"{u}}lah}}, \bibinfo {author} {\bibfnamefont {L.}~\bibnamefont {Marot}},
  \bibinfo {author} {\bibfnamefont {R.}~\bibnamefont {Steiner}}, \bibinfo
  {author} {\bibfnamefont {A.}~\bibnamefont {Romanyuk}}, \bibinfo {author}
  {\bibfnamefont {T.~A.}\ \bibnamefont {Jung}}, \bibinfo {author}
  {\bibfnamefont {A.}~\bibnamefont {W{\"{a}}ckerlin}}, \ and\ \bibinfo {author}
  {\bibfnamefont {E.}~\bibnamefont {Meyer}},\ }\href {10.1038/srep43369}
  {\bibfield  {journal} {\bibinfo  {journal} {{Scientific Reports}}\ }\textbf
  {\bibinfo {volume} {7}},\ \bibinfo {pages} {1} (\bibinfo {year}
  {2017})}\BibitemShut {NoStop}%
\bibitem [{\citenamefont {Craciun}\ \emph {et~al.}(1999)\citenamefont
  {Craciun}, \citenamefont {Howard}, \citenamefont {Lambers}, \citenamefont
  {Singh}, \citenamefont {Craciun},\ and\ \citenamefont
  {Perriere}}]{Craciun1999}%
  \BibitemOpen
  \bibfield  {author} {\bibinfo {author} {\bibfnamefont {V.}~\bibnamefont
  {Craciun}}, \bibinfo {author} {\bibfnamefont {J.}~\bibnamefont {Howard}},
  \bibinfo {author} {\bibfnamefont {E.}~\bibnamefont {Lambers}}, \bibinfo
  {author} {\bibfnamefont {R.}~\bibnamefont {Singh}}, \bibinfo {author}
  {\bibfnamefont {D.}~\bibnamefont {Craciun}}, \ and\ \bibinfo {author}
  {\bibfnamefont {J.}~\bibnamefont {Perriere}},\ }\href@noop {} {\bibfield
  {journal} {\bibinfo  {journal} {Applied Physics A}\ }\textbf {\bibinfo
  {volume} {69}},\ \bibinfo {pages} {S535} (\bibinfo {year}
  {1999})}\BibitemShut {NoStop}%
\bibitem [{\citenamefont {Barshilia}\ \emph {et~al.}(2012)\citenamefont
  {Barshilia}, \citenamefont {Chaudhary}, \citenamefont {Kumar},\ and\
  \citenamefont {Manikandanath}}]{388382:8582575}%
  \BibitemOpen
  \bibfield  {author} {\bibinfo {author} {\bibfnamefont {H.~C.}\ \bibnamefont
  {Barshilia}}, \bibinfo {author} {\bibfnamefont {A.}~\bibnamefont
  {Chaudhary}}, \bibinfo {author} {\bibfnamefont {P.}~\bibnamefont {Kumar}}, \
  and\ \bibinfo {author} {\bibfnamefont {N.~T.}\ \bibnamefont
  {Manikandanath}},\ }\href {10.3390/nano2010065} {\bibfield  {journal}
  {\bibinfo  {journal} {{Nanomaterials (Basel, Switzerland)}}\ }\textbf
  {\bibinfo {volume} {2}},\ \bibinfo {pages} {65} (\bibinfo {year}
  {2012})}\BibitemShut {NoStop}%
\bibitem [{\citenamefont {Mongstad}\ \emph
  {et~al.}(2011{\natexlab{b}})\citenamefont {Mongstad}, \citenamefont
  {Platzer-Björkman}, \citenamefont {Karazhanov}, \citenamefont {Holt},
  \citenamefont {Maehlen},\ and\ \citenamefont {Hauback}}]{MONGSTAD2011S812}%
  \BibitemOpen
  \bibfield  {author} {\bibinfo {author} {\bibfnamefont {T.}~\bibnamefont
  {Mongstad}}, \bibinfo {author} {\bibfnamefont {C.}~\bibnamefont
  {Platzer-Björkman}}, \bibinfo {author} {\bibfnamefont {S.}~\bibnamefont
  {Karazhanov}}, \bibinfo {author} {\bibfnamefont {A.}~\bibnamefont {Holt}},
  \bibinfo {author} {\bibfnamefont {J.}~\bibnamefont {Maehlen}}, \ and\
  \bibinfo {author} {\bibfnamefont {B.}~\bibnamefont {Hauback}},\ }\href@noop
  {} {\bibfield  {journal} {\bibinfo  {journal} {Journal of Alloys and
  Compounds}\ }\textbf {\bibinfo {volume} {509}},\ \bibinfo {pages} {S812 }
  (\bibinfo {year} {2011}{\natexlab{b}})}\BibitemShut {NoStop}%
\bibitem [{\citenamefont {Maehlen}\ \emph {et~al.}(2013)\citenamefont
  {Maehlen}, \citenamefont {Mongstad}, \citenamefont {You},\ and\ \citenamefont
  {Karazhanov}}]{MAEHLEN2013S119}%
  \BibitemOpen
  \bibfield  {author} {\bibinfo {author} {\bibfnamefont {J.~P.}\ \bibnamefont
  {Maehlen}}, \bibinfo {author} {\bibfnamefont {T.~T.}\ \bibnamefont
  {Mongstad}}, \bibinfo {author} {\bibfnamefont {C.~C.}\ \bibnamefont {You}}, \
  and\ \bibinfo {author} {\bibfnamefont {S.}~\bibnamefont {Karazhanov}},\
  }\href@noop {} {\bibfield  {journal} {\bibinfo  {journal} {Journal of Alloys
  and Compounds}\ }\textbf {\bibinfo {volume} {580}},\ \bibinfo {pages} {S119 }
  (\bibinfo {year} {2013})}\BibitemShut {NoStop}%
\bibitem [{\citenamefont {Pishtshev}\ \emph {et~al.}(2018)\citenamefont
  {Pishtshev}, \citenamefont {Strugovshchikov},\ and\ \citenamefont
  {Karazhanov}}]{388382:8582469}%
  \BibitemOpen
  \bibfield  {author} {\bibinfo {author} {\bibfnamefont {A.}~\bibnamefont
  {Pishtshev}}, \bibinfo {author} {\bibfnamefont {E.}~\bibnamefont
  {Strugovshchikov}}, \ and\ \bibinfo {author} {\bibfnamefont {S.}~\bibnamefont
  {Karazhanov}},\ }\href {\doibase 10.26434/chemrxiv.6950138.v1} {\bibfield
  {journal} {\bibinfo  {journal} {{Chemrxiv}}\ } (\bibinfo {year} {2018}),\
  10.26434/chemrxiv.6950138.v1}\BibitemShut {NoStop}%
\bibitem [{\citenamefont {Bersuker}(2013)}]{388382:8581607}%
  \BibitemOpen
  \bibfield  {author} {\bibinfo {author} {\bibfnamefont {I.~B.}\ \bibnamefont
  {Bersuker}},\ }\href {10.1021/cr300279n} {\bibfield  {journal} {\bibinfo
  {journal} {{Chem Rev}}\ }\textbf {\bibinfo {volume} {3}},\ \bibinfo {pages}
  {1351–1390} (\bibinfo {year} {2013})}\BibitemShut {NoStop}%
\end{thebibliography}%
\bibliographystyle{aip}
\bibliographystyle{apsrev4-1}

\end{document}